\documentclass[review,a4paper,12pt]{elsarticle}

\usepackage[title,titletoc,toc]{appendix}
\usepackage{graphicx}
\usepackage{color}
\usepackage[dvips,colorlinks,bookmarksopen,bookmarksnumbered,citecolor=blue,urlcolor=red]{hyperref}
\usepackage{amssymb,amsthm,amsmath}
\usepackage{bm}

\newcommand{\figref}[1]{Fig.~\ref{#1}}

\newcommand{\Equref}[1]{Eq.(\ref{#1})}
\newcommand{\equref}[1]{eq.(\ref{#1})}

\newcommand{\V}[1]{\mathbb{V}\left[#1\right]}
\newcommand{\E}[1]{\mathbb{E}\left[#1\right]}
\newcommand{\N}[1]{\mathcal{N}\left(#1\right)}
\newcommand{\Cov}[1]{{\rm Cov}\left(#1\right)}
\renewcommand{\d}{{\rm d}}

\title{Monte Carlo estimators of first- and total-orders Sobol' indices}

\author[rvt]{Ivano Azzini} 
\ead{Ivano.Azzini@ec.europa.eu}

\author[rvt]{Thierry A. Mara\corref{cor1}}
\ead{Thierry.Mara@ec.europa.eu}

\author[rvt]{Rossana Rosati}
\ead{Rossana.Rosati@ec.europa.eu}

\cortext[cor1]{Corresponding author}

\address[rvt]{European Commission, Joint Research Centre}

\begin{document}

\begin{frontmatter}
\begin{abstract}
This study compares the performances of two sampling-based strategies for the simultaneous estimation of the first- and total-orders variance-based sensitivity indices (a.k.a Sobol' indices). The first strategy was introduced by \cite{Saltelli02CPC} and is the current approach employed by practitioners. The second one was only recently introduced by the authors of the present article. They both rely on different estimators of first- and total-orders Sobol' indices. The asymptotic normal variances of the two sets of estimators are established and their accuracies are compared theoretically and numerically. The results show that the new strategy outperforms the current one.
\end{abstract}
\end{frontmatter}

\section{Background}
Uncertainty and sensitivity analysis is an essential ingredient of modelling \cite{Saltelli04BOOK}. It allows to point out the key uncertain assumptions (input factors that can be random variables or random fields) responsible for the uncertainty into the model outcome of interest. This is particularly relevant when models are used for decision-making. 

Assessing model output uncertainty requires several runs of the model. Monte Carlo simulations allow to carry out this task by sampling the input factors accordingly with their presumed joint probability distribution and propagating the sample (i.e. running the model) through the model response of interest. Sensitivity analysis (SA) can then be undertaken to identify the most relevant input factors. Depending on the method used, SA can be conducted directly from the Monte Carlo sample at hand (i.e., the one generated to assess model output uncertainty) or can require extra Monte Carlo simulations by following an appropriate sampling design.

The method to be used depends on the sensitivity indices (also called importance measures) that the analyst wants to compute. As recommended in \cite{Saltelli04BOOK} (see also \citep{Saltelli02JASA}), the sensitivity indices to assess should be related to the question that SA is called to answer to. The same authors enumerate several questions (called SA \emph{settings}) that can be addressed with the so-called variance-based sensitivity indices. In the sequel, we focus on the estimation of variance-based sensitivity indices, also called Sobol' indices (\citep{Sobol93MMCE}).

As eluded previously, a Monte Carlo sample is required to carry out uncertainty analysis (UA), that is, assessing the predictive uncertainty of the model outputs. In the sequel, we assume that there is only one scalar output denoted $y=f(\bm{x})$. The input factors are represented by a random vector of scalar variables $\bm{x}=(x_1,\dots,x_d)$ possibly grouped into two complementary vectors $(\bm{u},\bm{v})$. They are assumed independent of each others (for the case of dependent inputs, see for instance \citep{Mara15EMS}).

There exist several Sobol' indices called, first-order, (closed) second-order, and so forth. Of particular interest are the first- and total-orders Sobol' indices defined as follows,
\begin{equation}
\label{Eq:Si}
S_u = \frac{\V{\E{y\vert\bm{u}}}}{\V{y}}
\end{equation}
\begin{equation}
\label{Eq:STi}
ST_v = \frac{\E{\V{y\vert\bm{u}}}}{\V{y}}
\end{equation}
where, $\V{\cdot}$ stands for the unconditional variance operator (resp. $\V{\cdot\vert\cdot}$ the conditional variance) and $\E{\cdot}$ stands for the mathematical expectation (resp. $\E{\cdot\vert\cdot}$ the conditional expectation). We denote by $\bm{u}^{\sharp}$ the number of elements in $\bm{u}$.

\Equref{Eq:Si} is the first-order Sobol' index of the group of inputs $\bm{u}$ while \Equref{Eq:STi} is the total-order Sobol' index of $\bm{z}$. When $\bm{u}^{\sharp}=k>1$, $S_u$ is called the $k$-th order closed index and is often denoted $S^c_u$ (see \citep{Saltelli02CPC}). $S_u$ represents in percentage, the expected reduction in $\V{y}$ if the variables in $\bm{u}$ where fixed to their \emph{true} value. That is why the individual (i.e $\bm{u}^{\sharp}=1$) first-order sensitivity indices are to be estimated if the goal of the SA is to identify the input variable that would induce the largest reduction in variance if its value was known accurately. This SA setting is called \emph{factors prioritization}. Instead, if the goal is to identify the irrelevant inputs (called \emph{screening} or \emph{factors fixing setting}) then the individual total-order Sobol' indices are to be estimated. Indeed, we note that, $S_u+ST_v = 1$, which means that if $ST_v=0$, the variables in $\bm{v}$ do not contribute at all to the variance of $y$.

If the input-output relationship is smooth enough and $d$ is not too high, SA can be conducted after building a surrogate model from the input-output Monte Carlo sample used for UA (among others, \citep{Oakley04JRSS,Marrel09RESS,Buzzard11CCP,Blatman11JCP,Shao17CMAME}). By smooth we mean that $y$ is indefinitely derivable w.r.t. all the input factors and that the input-output relationship is not strongly non-linear. Then, Monte Carlo estimators are applied to the surrogate model to obtain the desired sensitivity indices. Monte Carlo estimators are rather computationally expensive, but they do not require any assumption on the input-output relationship but that the variance of $f(\bm{x})$ be computable. In the present work, we study the performances of two Monte Carlo estimators of Eqs.(\ref{Eq:Si}-\ref{Eq:STi}) that rely on two different sampling designs.

The paper is organised as follows: in Section \ref{Sec:Estimators} we introduce the two sampling strategies as well as their associated Monte Carlo estimators to compute both the first- and total-orders Sobol' indices. Their asymptotic normal variances, derived in the appendices, are also compared to each other. In Section \ref{Sec:Numerical}, the performances of the two estimators are compared through numerical exercises on notorious benchmark functions. The key results are summarized in Section \ref{Sec:Conclusion}.

\section{Monte Carlo estimators}
\label{Sec:Estimators}
\subsection{Integral approximation}
When Ilya M. Sobol' introduced for the first time the variance-based sensitivity indices in \cite{Sobol93MMCE}, he also proposed their Monte Carlo (MC) estimators. The latter rely on the fact that multidimensional integrals can be approximated by Monte Carlo samples as follows,
\begin{equation}
\label{Eq:MCE}
\int f(x_1,\cdots, x_d)p_x(\bm{x})\d\bm{x}\approx \frac{1}{N}\sum_{k=1}^{N}f(x_{k1},\cdots, x_{kd})
\end{equation}
where $\bm{x}\sim p_x$, meaning that $p_x$ is the joint probability density of $\bm{x}$ and $\bm{x}_k=(x_{k1},\cdots, x_{kd})$ is the $k$-th (out of $N$) MC draw of the input factors sampled w.r.t. $p_x$.

Let $(\bm{y}^A,\bm{y}^B,\bm{y}^{A_u},\bm{y}^{B_u})$ be four distinct model output samples whose $k$-th element for each of them is respectively defined as follows,
$$y_k^A = f(\bm{u}_{k}^{A},\bm{v}_{k}^{A}) = f(\bm{x}_{k}^A)$$
$$y_k^B = f(\bm{u}_{k}^{B},\bm{v}_{k}^{B}) = f(\bm{x}_{k}^B)$$
$$y_k^{A_u} = f(\bm{u}_{k}^{A},\bm{v}_{k}^{B}) = f(\bm{x}_{k}^{A_u})$$
$$y_k^{B_u} = f(\bm{u}_{k}^{B},\bm{v}_{k}^{A}) = f(\bm{x}_{k}^{B_u})$$
where $\bm{x}_k^A$ and $\bm{x}_k^B$ are two independent input vectors identically distributed, as well as $\bm{x}_k^{A_u}$ and $\bm{x}_k^{B_u}$. The $u$-values in vector $\bm{x}_k^{A_u}$  are identical to those in $\bm{x}_k^{A}$ while the $v$-values are those of $\bm{x}_k^{B}$.

\subsection{Current estimators}
The most popular sampling design to compute simultaneously first- and total-orders sensitivity indices was proposed by Saltelli \cite{Saltelli02CPC}. The latter requires three samples, namely $(\bm{y}^A,\bm{y}^B,\bm{y}^{A_u})$, to compute the sensitivity indices of $\bm{u}$. Their estimators are respectively defined as follows,
\begin{equation}
\label{Eq:Si_SS}
\hat{S}^{SS}_u = \frac{2\sum_{k=1}^N y_k^A\left(y_k^{A_u}-y_k^B\right)}{\sum_{k=1}^N\left(y_k^A-y_k^B\right)^2}
\end{equation}
\begin{equation}
\label{Eq:STi_SJ}
\hat{ST}^{SJ}_u = \frac{\frac{1}{N}\sum_{k=1}^N \left(y_k^{A_u}-y_k^B\right)^2}{\frac{1}{N}\sum_{k=1}^{N}\left(y_k^A-y_k^B\right)^2}
\end{equation}

Note that there exist various versions of the estimators, especially regarding the denominator. We find it convenient to formulate it in this way because it highlights the symmetry between $(\bm{y}^A,\bm{y}^B)$ in the denominator. \Equref{Eq:Si_SS} is known to provide accurate estimate of small first-order sensitivity indices \cite{Sobol07RESS} while \Equref{Eq:STi_SJ} is called the Sobol-Jansen estimator and was introduced in \cite{Jansen99CPC}. The performance of an estimator is characterized by its \emph{bias} and its \emph{variance}. MC estimators such as \Equref{Eq:MCE} are unbiased. In terms of variance, the estimators in Eqs.(\ref{Eq:Si_SS}-\ref{Eq:STi_SJ}) differ quite much.

More importantly, although in theory $ST_u\geq S_u$, the previous estimators do no satisfy this criterion. Indeed, we note that,
\begin{equation}
\label{Eq:Difference1}
\sum_{k=1}^{N}\left(y_k^A-y_k^B\right)^2\left(\hat{ST}^{SJ}_u-\hat{S}^{SS}_u\right) = \sum_{k=1}^N \left(y_k^{A_u}-y_k^B\right)^2-2y_k^A\left(y_k^{A_u}-y_k^B\right)
\end{equation}
which, because $-2y_k^A\left(y_k^{A_u}-y_k^B\right)$ can be either positive or negative, does not ensure that $\hat{ST}^{SJ}_u\geq \hat{S}^{SS}_u$.

These observations advocate for a more symmetrical and coherent estimator for the first-order sensitivity index. This is the subject of the next subsection.

\subsection{New estimators}
By noticing that the denominator of \Equref{Eq:STi_SJ} converges towards $2\V{y}$, that is,
$$\lim_{N\rightarrow\infty}\frac{1}{N}\sum_{k=1}^{N}\left(y_k^A-y_k^B\right)^2=\lim_{N\rightarrow\infty}\sum_{k=1}^{N}\frac{1}{N}\left(y_k^{A_u}-y_k^{B_u}\right)^2=2\V{y}$$
and that the numerator is such that,
$$\lim_{N\rightarrow\infty}\frac{1}{N}\sum_{k=1}^N \left(y_k^B-y_k^{A_u}\right)^2=\lim_{N\rightarrow\infty}\frac{1}{N}\sum_{k=1}^N \left(y_k^A-y_k^{B_u}\right)^2=2\E{\V{y\vert\bm{v}}}$$
the following symmetrical estimator for the total-order sensitivity index can be derived,
\begin{equation}
\label{Eq:STi_IA}
\hat{ST}^{IA}_u = \frac{\sum_{k=1}^N \left(y_k^B-y_k^{A_u}\right)^2+\left(y_k^A-y_k^{B_u}\right)^2}{\sum_{k=1}^{N}\left(y_k^A-y_k^B\right)^2+\left(y_k^{A_u}-y_k^{B_u}\right)^2}.
\end{equation}
This is because, as already mentioned, $\bm{x}_k^A$ and $\bm{x}_k^B$ are two independent input vectors identically distributed, as well as $\bm{x}_k^{A_u}$ and $\bm{x}_k^{B_u}$. Notice the perfect symmetry of the formula which remain unchanged by exchanging the superscripts referring to $B$ with $A$. Incidentally, the superscript IA stands indifferently for Improved Algorithm and Ivano Azzini the first author of this article who guessed this formula.

The new first-order estimator can then be inferred as,
\begin{equation}
\label{Eq:Si_IA}
\hat{S}^{IA}_u = \frac{2\sum_{k=1}^N \left(y_k^{A_u}-y_k^B\right)\left(y_k^{A}-y_k^{B_u}\right)}{\sum_{k=1}^{N}\left(y_k^A-y_k^B\right)^2+\left(y_k^{A_u}-y_k^{B_u}\right)^2}
\end{equation}
Furthermore, we easily prove that $\hat{ST}^{IA}_u\geq \hat{S}^{IA}_u$.

\begin{proof}
Interchanging $(\bm{y}^{A_u},\bm{y}^{B_u})$ in \Equref{Eq:STi_IA} only changes the numerator and provides the estimator for $\hat{ST}^{IA}_v$. Therefore, the first-order sensitivity index $S_u$ is estimated as follows,
\begin{equation*}
\hat{S}^{IA}_u = 1 - \hat{ST}^{IA}_v = 1 - \frac{\sum_{k=1}^N \left(y_k^B-y_k^{B_u}\right)^2+\left(y_k^A-y_k^{A_u}\right)^2}{\sum_{k=1}^{N}\left(y_k^A-y_k^B\right)^2+\left(y_k^{A_u}-y_k^{B_u}\right)^2}
\end{equation*}
which after some developments yields \Equref{Eq:Si_IA}.

Besides,
\begin{equation*}
\hat{ST}^{IA}_u-\hat{S}^{IA}_u = \frac{\sum_{k=1}^N \left(y_k^B-y_k^{A_u}\right)^2+\left(y_k^A-y_k^{B_u}\right)^2 - 2\left(y_k^{A_u}-y_k^B\right)\left(y_k^{A}-y_k^{B_u}\right)}{\left(\sum_{k=1}^{N}\left(y_k^A-y_k^B\right)^2+\left(y_k^{A_u}-y_k^{B_u}\right)^2\right)}
\end{equation*}
\begin{equation}
\label{Eq:Difference2}
\hat{ST}^{IA}_u-\hat{S}^{IA}_u = \frac{\sum_{k=1}^N \left(y_k^B-y_k^{A_u}+y_k^A-y_k^{B_u}\right)^2}{\sum_{k=1}^{N}\left(y_k^A-y_k^B\right)^2+\left(y_k^{A_u}-y_k^{B_u}\right)^2}\geq 0
\end{equation}
\end{proof}
\Equref{Eq:Difference2} also shows that $\hat{ST}^{IA}_u=\hat{S}^{IA}_u$ if and only if $f(\bm{x})$ is additive with respect to $\bm{u}$, that is, $ST_u=S_u$. In effect, we can write in this case,
$$y=f(\bm{u},\bm{v})=f_0+f_u(\bm{u})+f_v(\bm{v})$$
and it is straightforward to prove that the numerator of \Equref{Eq:Difference2} equals zero, and so, whatever the sample size $N$. 

\subsection{Estimators' variances}
\label{Sec:Discussion}
In the Appendices \ref{Sec:A1} and \ref{Sec:A2}, we establish the variances of the estimators discussed in the present paper under the asymptotic normality assumption \cite{Waart2000BOOK,Janon14ESAIM}. They respectively read as follows,
\begin{eqnarray}
\label{Eq:Var_SS}
\sigma_{SS}^2 &=& \frac{\V{2y^A\left(y^{A_u}-y^{B}\right) - S_u\left(y^A-y^B\right)^2}}{4N\V{y}^2}\\
\label{Eq:Var_SJ}
\tau^2_{SJ} &=& \frac{\V{\left(y^{A_u}-y^B\right)^2-ST_{u}\left(y^A-y^B\right)^2}}{4N\V{y}^2}
\end{eqnarray}
 and,
\begin{eqnarray}
\label{Eq:Var_IA}
\sigma_{IA}^2 &=& \frac{\V{2\left(y^A-y^{B_u}\right)\left(y^{A_u}-y^B\right) -S_u\left(\left(y^A-y^B\right)^2+\left(y^{A_u}-y^{B_u}\right)^2\right)}}{2(2\times 4N\V{y}^2)}\\
\label{Eq:Var_IA_Total}
\tau_{IA}^2 &=& \frac{\V{\left(y^A-y^{B_u}\right)^2+\left(y^B-y^{A_u}\right)^2 -ST_u\left(\left(y^A-y^B\right)^2+\left(y^{A_u}-y^{B_u}\right)^2\right)}}{2(2\times 4N\V{y}^2)}
\end{eqnarray}
First of all, we notice that the current estimators Eqs.(\ref{Eq:Si_SS}-\ref{Eq:STi_SJ}) require $N(d+2)$ model calls to estimate the overall set of first- and total-orders Sobol' indices while Eqs.(\ref{Eq:Si_IA}-\ref{Eq:STi_IA}) require $2N(d+1)$. Thus, the new estimators require approximately twice more samples. To ensure a fair comparison, we take into account this difference by highlighting this factor 2 in the denominators of Eqs.(\ref{Eq:Var_IA}-\ref{Eq:Var_IA_Total}) as compared to Eqs.(\ref{Eq:Var_SS}-\ref{Eq:Var_SJ}).

It can be qualitatively guessed that $\tau^2_{SJ}\leq \sigma_{SS}^2$. Indeed, we have (according to \citep{Sobol93MMCE}),
\begin{equation*}
y = f(\bm{u},\bm{v}) = f_0+f_u(\bm{u})+f_v(\bm{v})+f_{u,v}(\bm{u},\bm{v})
\end{equation*}
This implies that,
\begin{eqnarray*}
\left(y^{A_u}-y^B\right) &=& -f_u(\bm{u}^B)+f_u(\bm{u}^A)-f_{u,v}(\bm{u}^B,\bm{v}^B)+f_{u,v}(\bm{u}^A,\bm{v}^B)\\
\left(y^{A}-y^{B_u}\right) &=& -f_u(\bm{u}^B)+f_u(\bm{u}^A)-f_{u,v}(\bm{u}^B,\bm{v}^A)+f_{u,v}(\bm{u}^A,\bm{v}^A)
\end{eqnarray*}
Therefore, the variance of $\left(y^{B_u}-y^{A}\right)^2$ is expected to be smaller than $2y^A$ $\left(y^{A_u}-y^{B}\right)$ because the former does not contain neither $f_0$, nor $f_v$ contrarily to the latter with $y^A$. What is worse, the estimator (\ref{Eq:Si_SS}) may perform very poorly for high values of $f_0$. Besides, we note that $\left(y^B-y^{A_u}\right)\left(y^{B_u}-y^{A}\right) \sim \left(y^{B_u}-y^{A}\right)^2$ which indicates that $\sigma^2_{IA}\leq \sigma_{SS}^2$. Nevertheless, it is less obvious to infer whether $\tau^2_{SJ}$ is higher or lower than $\tau_{IA}^2$. Therefore, this is investigated through numerical simulations in the next section.

 \section{Numerical examples}
 \label{Sec:Numerical}
It is worth noting that the current estimators Eqs.(\ref{Eq:Si_SS}-\ref{Eq:STi_SJ}) require $N(d+2)$ model calls to estimate the overall set of first- and total-orders Sobol' indices while Eqs.(\ref{Eq:Si_IA}-\ref{Eq:STi_IA}) require $2N(d+1)$. To ensure a fair comparison, we set the sample size of the new estimators to half the one of the current estimators. In this way, the computational cost is $2N(d+1)$ for the former and $2N(d+2)$ for the latter. This means that when we write that a sample of size $N$ is used, this refers to the actual size of the samples for the new estimators while the sample size is $2N$ for the current estimators Eqs.(\ref{Eq:Si_SS}-\ref{Eq:STi_SJ}).

 \subsection{The Ishigami function}
 Let us consider the following three-dimensional function,
 \begin{equation}
 f(x_1,x_2,x_3) = f_0 + \sin x_1+7\sin^2 x_2+0.1x_3^4\sin x_1
 \end{equation}
 where the input variables are independently an uniformly distributed over $(-\pi,\pi)^3$. As compared to the original Ishigami function, we introduce a constant parameter $f_0$ which has no impact on the variance of the function. This simple function for which the exact Sobol' indices are known has the following features: $x_1$ and $x_3$ interact strongly while $x_2$ is \emph{additively} influential, that is, $S_2=ST_2\simeq 0.44$. This allows to check whether, as previously guessed, we find $\hat{S}_2^{IA}=\hat{ST}_2^{IA}$. In this exercise, we numerically compare the performances of Eqs.(\ref{Eq:Si_SS}-\ref{Eq:STi_SJ}) with Eqs.(\ref{Eq:Si_IA}-\ref{Eq:STi_IA}). For this purpose, we set $N=64$ and we assess 100 replicate estimates of the first- and total-orders Sobol' indices with the estimators discussed in this paper.
 
\subsubsection{Case 1: $f_0=0$}
\begin{figure}[htbp]
	\centering
		\includegraphics[scale=0.75]{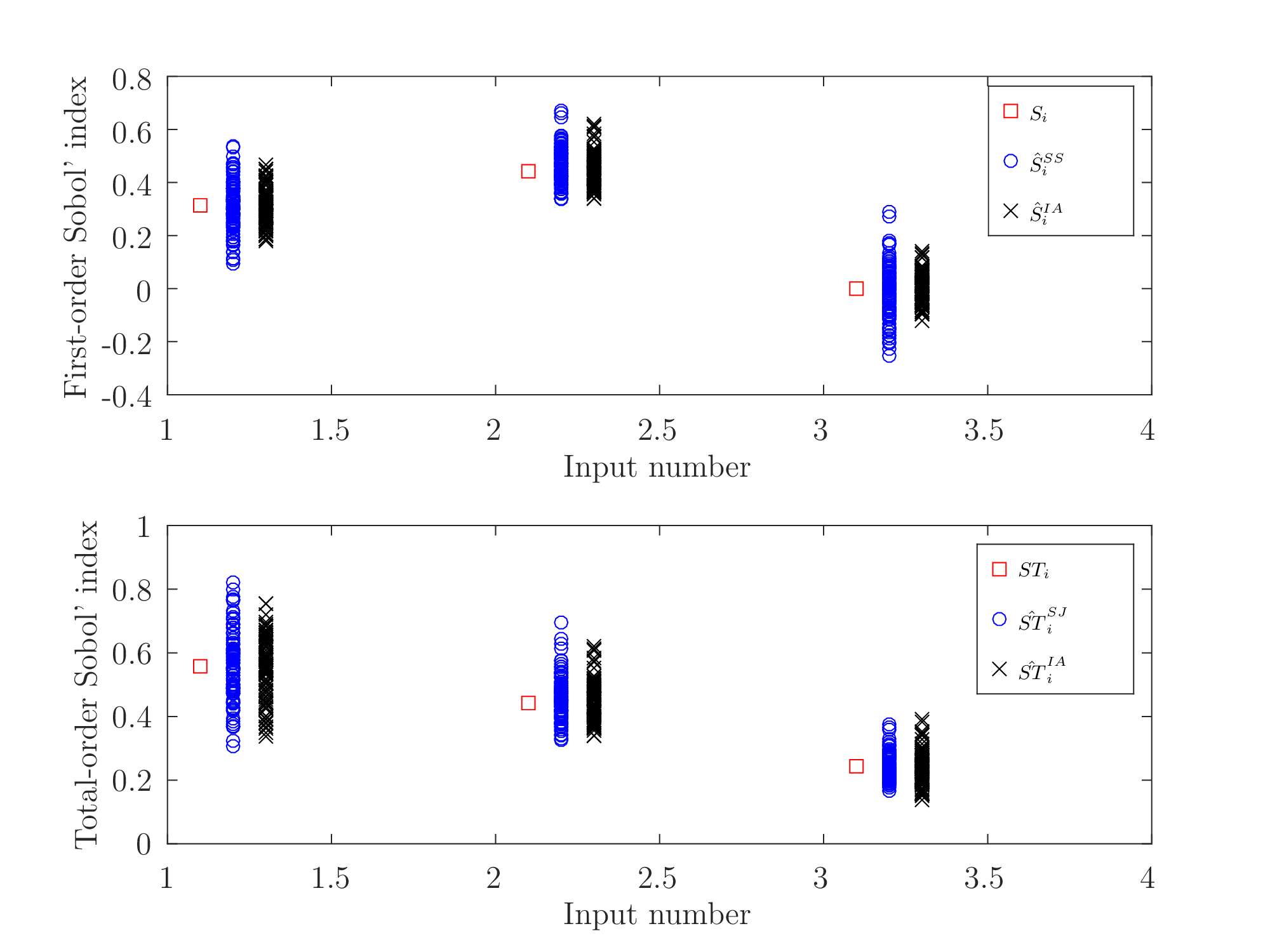}
	\caption{One hundred lhs-replicates of the first- and total-orders Sobol' indices (resp. at the top and the bottom) with the current and new estimators for the classical Ishigami function. For fair comparison, the sample size is $128$ for the current estimators and $64$ for the new ones.}
	\label{Fig1}
\end{figure}

We use the latin hypercube sampler (lhs) and first set $f_0=0$. The results are depicted in \figref{Fig1} which clearly shows that, as far as the first-order Sobol' indices are concerned, the new estimator \Equref{Eq:Si_IA} provides more robust estimates than \Equref{Eq:Si_SS}; thus confirming our comments in \S~\ref{Sec:Discussion}. Notably, $\hat{S}_3^{IA}$ the estimated first-order Sobol' index of $x_3$ can be smaller than zero which is not consistent with the theory (Sobol' indices shall be within [0,1]). This is due to its interaction with $x_1$. The new total-order estimator \equref{Eq:STi_IA} has slightly lower variances for $ST_1$ and $ST_2$ than \equref{Eq:STi_SJ} and conversely for $ST_3$.

\begin{figure}[htbp]
	\centering
		\includegraphics[scale=0.75]{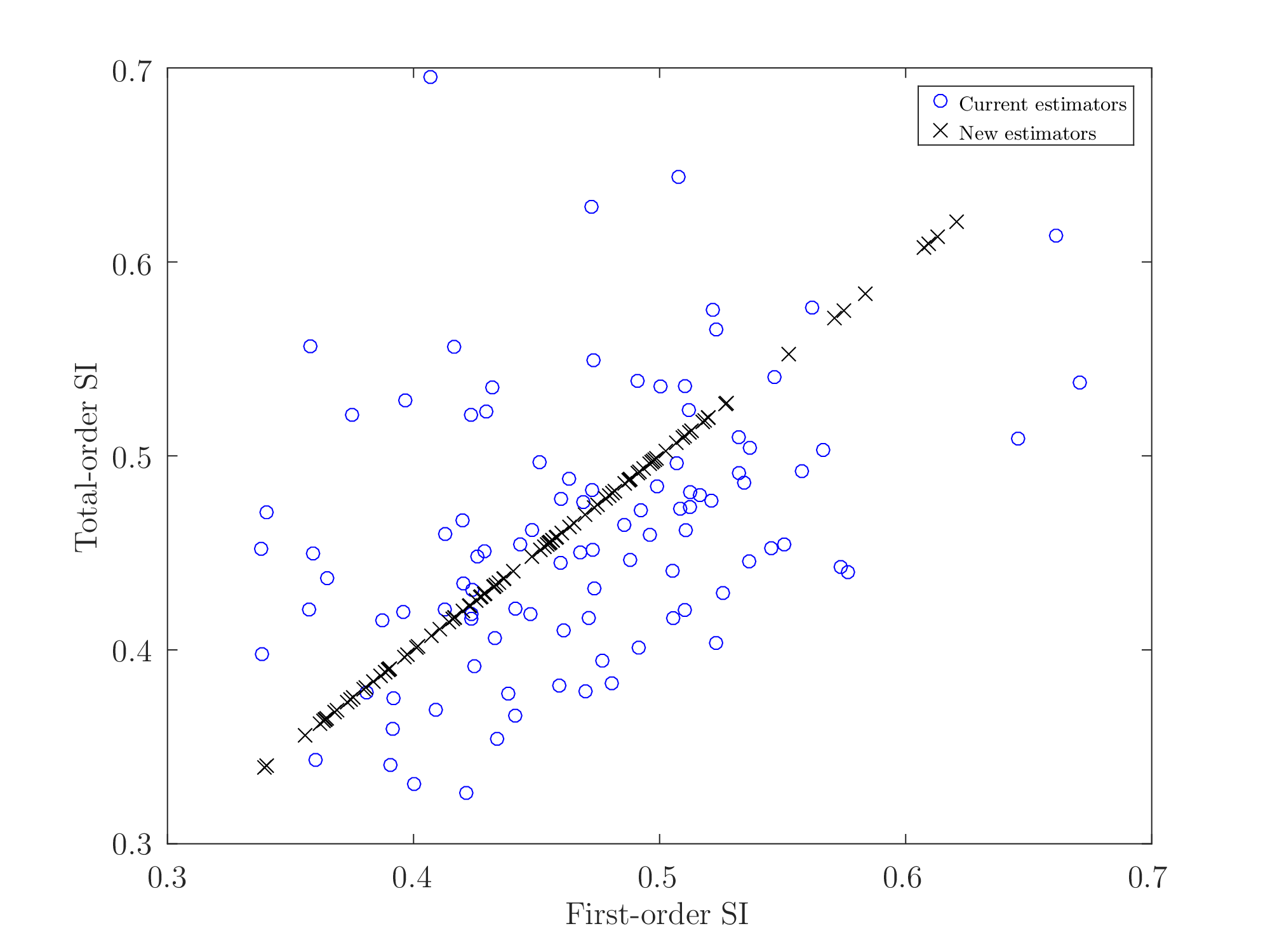}
	\caption{First- versus total-orders Sobol' indices of $x_2$ obtained with the current and new estimators for one hundred different lhs-replicates. The new estimators provide equal indices as $x_2$ does not interact with the other variables.}
	\label{Fig2}
\end{figure}

\figref{Fig2} depicts $\hat{S}_2$ versus $\hat{ST}_2$ for both couples of estimators (the current and new ones). We can see that $(\hat{S}_2^{IA},\hat{ST}_2^{IA})$ spreads along the line $\hat{S}_2^{IA}=\hat{ST}_2^{IA}$ contrarily to $(\hat{S}_2^{SS},\hat{ST}_2^{SJ})$. This is also in accordance with our findings in \S~\ref{Sec:Discussion} that $\hat{S}_i^{IA}=\hat{ST}_i^{IA}$ if  $x_i$ does not interact with the other variables. This is not the case with $(\hat{S}_2^{SS},\hat{ST}_2^{SJ})$. Actually for some replicates, we even find $\hat{S}_2^{SS}>\hat{ST}_2^{SJ}$ which is not consistent at all with the definition of first- and total-orders Sobol' indices. We stress that $\hat{S}_i^{IA}=\hat{ST}_i^{IA}$, when $x_i$ has only an additive effect on the response, is independent of the sample size $N$. This information can be obtained even at very low sample sizes (say $N\sim 10$).
 
 \subsubsection{Case 2: $f_0=100$}
 This case illustrates the sensitivity of the current first-order estimator to model responses with high expected value as compared with the total variance. We set $f_0=100$ keeping in mind that the Ishigami function has a total variance approximately equal to $V_y=13.84$. One hundred lhs-replicates of size $N=64$ (which means 128 for the current estimators) are employed. 
 
\begin{figure}[htbp]
	\centering
		\includegraphics[scale=0.75]{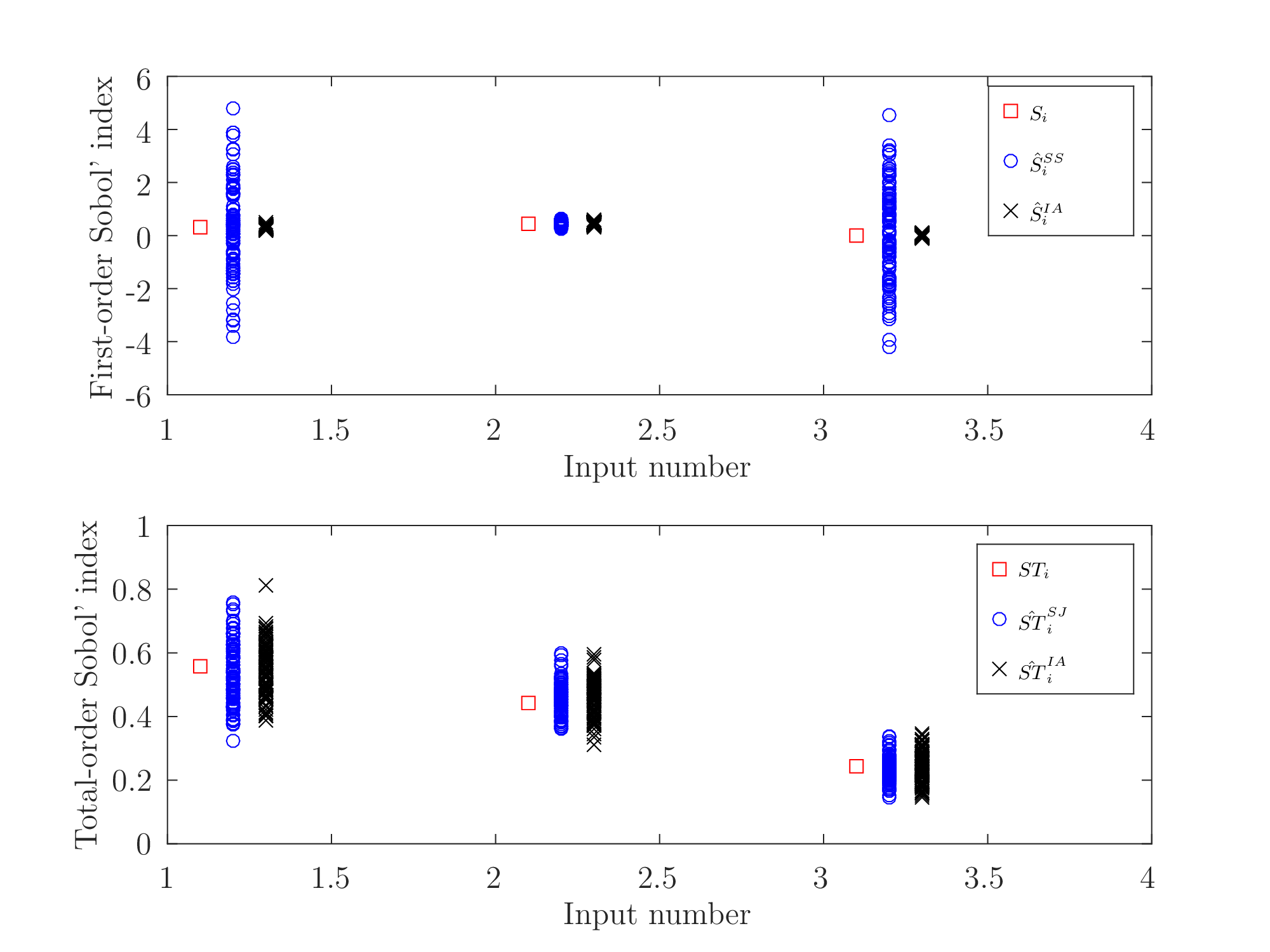}
	\caption{One hundred lhs-replicates of the first- and total-orders Sobol' indices (resp. at the top and the bottom) with the current and new estimators for the modified Ishigami function ($f_0=100$). In this case, the current estimator for first-order Sobol' index performs poorly (top).}
	\label{Fig3}
\end{figure}

The results are displayed in \figref{Fig3}. They show that while the shift in the Ishigami function has no impact on the estimators of the total-order estimators and on the new first-order estimator (namely, \Equref{Eq:Si_IA}), it significantly deteriorates the performance of the current first-order estimator (\Equref{Eq:Si_SS}) when the variables highly interact with each other. Indeed, on the top of \figref{Fig3} we can notice that $\hat{S}_2^{SS}$ is not affected. This result is in line with our comments in Section \ref{Sec:Discussion}.

Regarding the performance of the total-order estimators, it is not obvious to guess which one is better. A glance at the plot on the bottom of \figref{Fig3} reveals that the new estimator has lower variance for $ST_3$ and higher or equal variances for the two others. One might conclude that the new total-order estimator is more accurate for high total-order Sobol' indices. We investigate this hypothesis further in the next numerical exercise.

\subsection{The Sobol' function}
In this exercise, we study the performance of the two estimators of total-order Sobol' index. Specifically, we investigate whether the variance of the new estimator is always smaller than the current one or if it depends on the value of $ST_i$. For this purpose, we consider a ten-dimensional function whose total-order Sobol' indices of the input variables spread uniformly over $(0,1)$. Hence, we consider the Sobol' g-function defined as follows,
\begin{equation*}
f(\bm{x})=\prod_{i=1}^{10}\frac{\vert 4x_i-2\vert+a_i}{a_i+1}
\end{equation*}
where $x_i\sim\mathcal{U}(0,1)$ for all $i=1,\dots,10$ and the coefficients are chosen as follows:
$\bm{a}=\left(-1.13,-1.24,-1.33,-1.42,-1.52,-1.64,-1.79,-2.00,-2.37,1.52\right)$. This choice approximately yields the following total-order Sobol' indices, $\left(0.95,0.85,\dots,0.15,0.05\right)$. Thus $x_1$ has the highest total-order effect and $x_{10}$ the lowest.

\begin{figure}[htbp]
	\centering
		\includegraphics[scale=0.75]{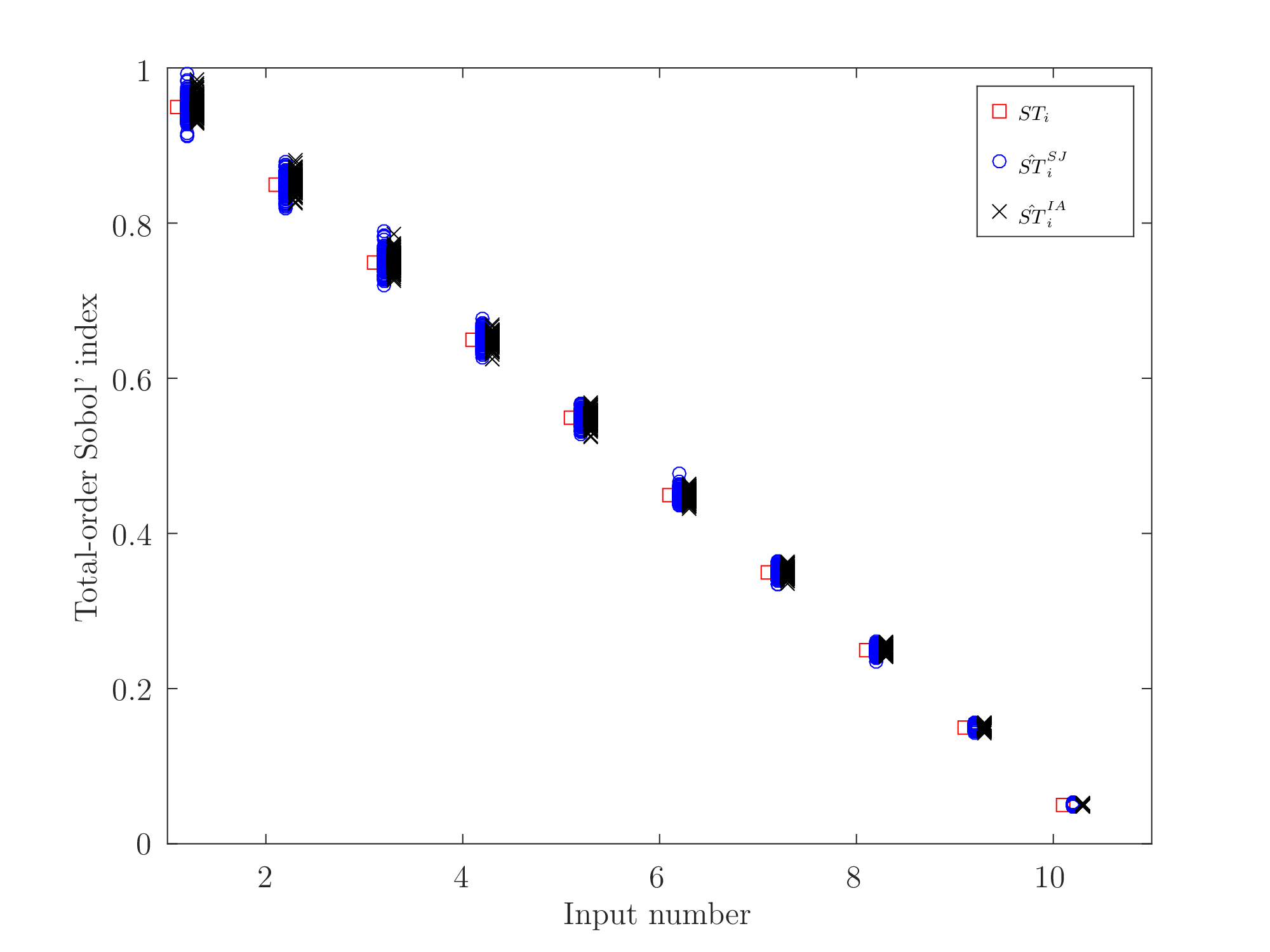}
	\caption{One hundred lhs-replicates of the total-orders Sobol' indices with the current and new estimators for the classical g-function. For fair comparison, the sample size is $2^{20}$ for the current estimators and $2^{21}$ for the new ones.}
	\label{Fig4}
\end{figure}

The numerical setting is as follows: we compute one hundred lhs-replicate estimates of the total-order sensitivity indices. Samples of size $N=2^{20}$ is employed ($2^{21}$ for the current estimator). For each estimate, the asymptotic normal variances Eqs.(\ref{Eq:Var_SJ}-\ref{Eq:Var_IA_Total}) are evaluated by replacing the exact Sobol' index (i.e. $ST_i$) and total variance (i.e., $\V{y}$) by their estimated value. The lhs-replicates provide also the empirical variances which can be confronted to the asymptotic normal variances. The one hundred estimates are depicted in \figref{Fig4} with the exact total-order Sobol' indices. The estimated Sobol' indices are very accurate and their range of variation does not overlap.

\begin{figure}[htbp]
	\centering
		\includegraphics[scale=0.75]{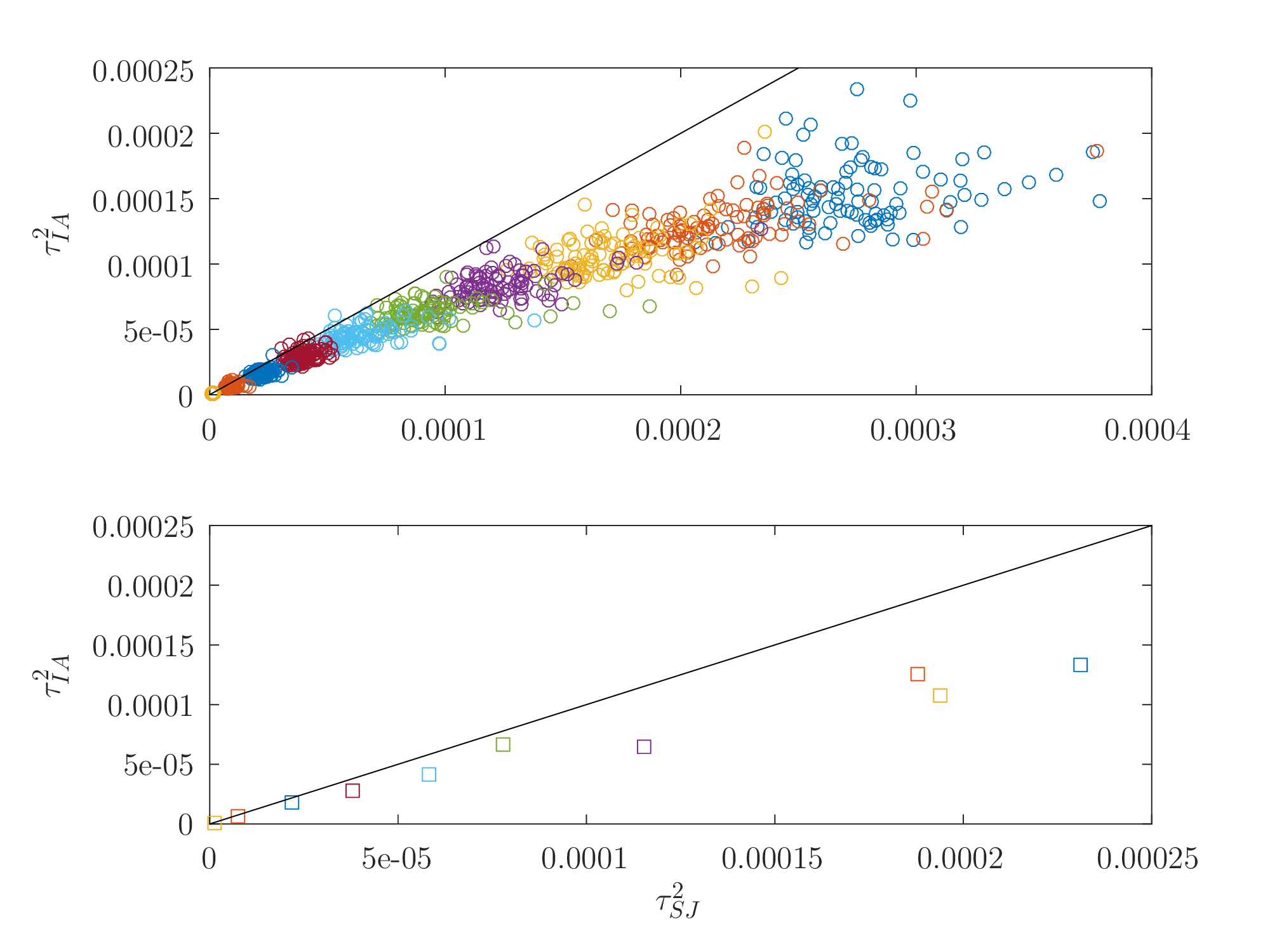}
	\caption{Estimated variances of the total-order SI estimators. On the top, by using the asymptotic normal variance formulas. On the bottom by evaluating the variances of the one hundred lhs-replicates. The continuous lines represent $\tau^2_{IA}=\tau^2_{SJ}$.}
	\label{Fig5}
\end{figure}

On the top of \figref{Fig5}, we represent the estimated variance of the new estimator (namely, $\tau_{IA}^2$) versus the variance of the current estimator ($\tau_{SJ}^2$). Because there are one hundred replicates of the sensitivity indices, for each sensitivity index $ST_i$, $i=1,\dots,10$, we have one hundred estimates of the asymptotic normal variances. They are depicted in different coloured circles in the top plot. On the bottom of \figref{Fig5}, we represent the empirical estimated variances obtained by computing directly the variance of the one hundred lhs-replicates of each total-order Sobol' index. First, we can note that while the $y$-axes of the two plots (bottom and top) have the same ranges, the ranges of $x$-axes are rather different (by virtually a factor of two). This indicates that \Equref{Eq:Var_IA_Total} is a good proxy of the empirical variance for the function under study unlike \Equref{Eq:Var_SJ} which seems to overestimate the \emph{true} estimator's variance.

The continuous line in \figref{Fig5} represents $\tau_{IA}^2=\tau_{SJ}^2$. The scatter plots located below this line means that $\tau_{IA}^2<\tau_{SJ}^2$. We observe that the scatter plots associated with the highest sensitivity indices (namely, from $ST_1$ to $ST_4$) are clearly below this lines either for the asymptotic normal variances (top) or the empirical variances (bottom). This confirms that, likewise the Ishigami function, the new estimator \Equref{Eq:STi_IA} is more accurate than \Equref{Eq:STi_SJ} at least for high sensitivity indices (say $ST_i>0.55$). Of course, this inference has been obtained numerically and extrapolation should be undertaken with caution.

\section{Conclusion}
\label{Sec:Conclusion}
We have introduced and studied the properties of two symmetrical MC estimators for first- and total-orders Sobol' indices respectively. It takes $2N(d+1)$ model calls to assess the overall set of indices with the associated sampling strategy. The new estimators possess interesting features. One of these features is that the estimated first-order index is always smaller than or equal to the total-order Sobol' index (unlike the current estimators mostly in use by practitioners).  By analysing their asymptotic normal variances and by conducting numerical exercises, we have shown that the new sampling strategy and its associated estimators perform better than the current estimator originally introduced in \cite{Saltelli02CPC}. Hence, we recommend the use of the IA-estimators to compute variance-based sensitivity indices with Monte Carlo integral approximation.

 \begin{appendices} 
 \section{Asymptotic normality of $\hat{S}^{SS}_u$ and $\hat{ST}^{SJ}_u$}
 \label{Sec:A1}
 The law of large numbers ensures that the estimator $\hat{S}^{SS}_u$ in \Equref{Eq:Si_SS} is consistent, that is,
 $$\lim_{N\rightarrow\infty}\hat{S}^{SS}_u=S_{u}$$ almost surely.
 
We denote by $\hat{S}^{SS}_u(N)$ the estimator for a sample size $N$. In the sequel, we follow the steps of \cite{Janon14ESAIM} to establish that the asymptotic normality of this estimator is,
 \begin{eqnarray}
\lim_{N\rightarrow\infty} \left(\hat{S}^{SS}_u(N)-S_{u}\right) &\sim & \N{0,\sigma^2_{SS}}
 \end{eqnarray}
with $\sigma^2_{SS}$ defined by \Equref{Eq:Var_SS}.

\begin{proof}
We set,
$$(\alpha_k,\beta_k)=\left(2y_k^A\left(y_k^{A_u}-y_k^B\right),\left(y_k^A-y_k^B\right)^2\right)$$ We also denote the associated random vector,$$(\alpha,\beta)=\left(2y^A\left(y^{A_u}-y^B\right),\left(y^A-y^B\right)^2\right)$$ since their statistics do not depend on $k$.

We then have,
$$(\bar{\alpha},\bar{\beta})=\lim_{N\rightarrow\infty}\frac{1}{N}\sum_{k=1}^N(\alpha_k,\beta_k)=\left(2S_{u}\V{y},2\V{y}\right)$$
and from \Equref{Eq:Si_SS} we can write,
\begin{equation*}
S_u = \phi(\bar{\alpha},\bar{\beta})=\frac{\bar{\alpha}}{\bar{\beta}}
\end{equation*}

The so-called Delta method \cite{Waart2000BOOK} allows for evaluating the variance of the estimator as follows,
\begin{eqnarray*}
\sigma_{SS}^2 = \frac{1}{N}g\Gamma g^t, & g = \nabla\phi(\bar{\alpha},\bar{\beta})
\end{eqnarray*}
with
\begin{equation*}
\Gamma = \left[\begin{matrix}
\V{\alpha} & \Cov{\alpha,\beta} \\
\Cov{\alpha,\beta} & \V{\beta}
\end{matrix}\right]
\end{equation*}
We find that,
\begin{eqnarray*}
g(\alpha,\beta) &=& \left(1/\beta, -\alpha/\beta^2\right)\\
\Leftrightarrow g(\bar{\alpha},\bar{\beta}) &=& \left(1/2\V{y}, -S_u/2\V{y}\right)
\end{eqnarray*}
by accounting for the definition of $(\bar{\alpha},\bar{\beta})$ above.

Therefore, we find that the variance of this estimator is,
\begin{equation*}
4N\V{y}^2\sigma_{SS}^2 = \V{\alpha}-2S_u\Cov{\alpha,\beta}+ S_u^2\V{\beta}
\end{equation*}
which can be rearranged as follows,
\begin{equation}
\label{Eq:alphabeta}
4N\V{y}^2\sigma_{SS}^2 = \V{\alpha-S_u\beta}
\end{equation}
Replacing $(\alpha,\beta)$ by their expression provides the announced result.
\end{proof}
Moreover, by noticing that in \Equref{Eq:alphabeta} $\alpha$ is the numerator of \Equref{Eq:Si_SS} and $\beta$ the denominator, it is straightforward to demonstrate that the variance of estimator \eqref{Eq:STi_SJ} is \Equref{Eq:Var_SJ}. This is merely established by setting $\alpha_k=\left(y_k^{A_u}-y_k^B\right)^2$, $\beta_k$ remaining unchanged.

 \section{Asymptotic normality of $\hat{S}^{IA}_u$ and $\hat{ST}^{IA}_u$}
 \label{Sec:A2}
 In the same way, it can be established that the asymptotic normality of $\hat{S}^{IA}_u$ is,
 \begin{equation}
\lim_{N\rightarrow\infty} \left(\hat{S}^{IA}_u(N)-S_{u}\right)\sim \N{0,\sigma^2_{IA}}
 \end{equation}
with $\sigma^2_{IA}$ given by \Equref{Eq:Var_IA}.

\begin{proof}
From \Equref{Eq:Si_IA} we can write,
\begin{equation*}
S_u = \phi(\bar{\alpha},\bar{\beta},\bar{\gamma})=\frac{\bar{\alpha}}{\bar{\beta}+\bar{\gamma}}
\end{equation*}
with,
$$(\alpha_k,\beta_k,\gamma_k)=\left(2\left(y_k^{B_u}-y_k^A\right)\left(y_k^B-y_k^{A_u}\right),\left(y_k^A-y_k^B\right)^2,\left(y_k^{A_u}-y_k^{B_u}\right)^2\right)$$
which yields,
$$(\bar{\alpha},\bar{\beta},\bar{\gamma})=\lim_{N\rightarrow\infty}\frac{1}{N}\sum_{k=1}^N(\alpha_k,\beta_k,\gamma_k)=\left(4S_{u}\V{y},2\V{y},2\V{y}\right)$$

We also denote the associated random vector,$$(\alpha,\beta,\gamma)=\left(2\left(y^{B_u}-y^A\right)\left(y^B-y^{A_u}\right),\left(y^A-y^B\right)^2,\left(y^{A_u}-y^{B_u}\right)^2\right)$$ since their statistics do not depend on $k$.

The so-called Delta method \cite{Waart2000BOOK} yields,
\begin{eqnarray*}
\sigma_{IA}^2 = \frac{1}{N}g\Gamma g^t, & g = \nabla\phi(\bar{\alpha},\bar{\beta},\bar{\gamma})
\end{eqnarray*}
with
\begin{equation*}
\Gamma = \left[\begin{matrix}
\V{\alpha} & \Cov{\alpha,\beta} & \Cov{\alpha,\gamma}\\
\Cov{\alpha,\beta} & \V{\beta} & \Cov{\beta,\gamma}\\
\Cov{\alpha,\gamma} & \Cov{\beta,\gamma} & \V{\gamma}
\end{matrix}\right]
\end{equation*}
We find that,
\begin{eqnarray*}
g(\alpha,\beta,\gamma) &=& \left(1/(\beta+\gamma), -\alpha/(\beta+\gamma)^2, -\alpha/(\beta+\gamma)^2)\right)\\
\Leftrightarrow g(\bar{\alpha},\bar{\beta},\bar{\gamma}) &=& \left(1/4\V{y}, -S_u/4\V{y}, -S_u/4\V{y}\right)
\end{eqnarray*}
by accounting for the definition of $(\bar{\alpha},\bar{\beta},\bar{\gamma})$ above.

Therefore, we find that the variance of our estimator is,
\begin{eqnarray*}
\begin{split}
16N\V{y}^2\sigma_{IA}^2 =& \V{\alpha}-2S_u\left[\Cov{\alpha,\beta}+\Cov{\alpha,\gamma}\right]+\\
& S_u^2\left[\underset{\V{\beta+\gamma}}{\underbrace{\V{\beta}+2\Cov{\beta,\gamma}+\V{\gamma}}}\right]
\end{split}
\end{eqnarray*}
which can be rearranged as follows,
\begin{equation*}
16N\V{y}^2\sigma_{IA}^2 = \V{\alpha}-2\Cov{\alpha,S_u\left(\beta+\gamma\right)}+ \V{S_u\left(\beta+\gamma\right)}
\end{equation*}
to finally give,
\begin{equation*}
\sigma_{IA}^2 = \frac{\V{\alpha-S_u\left(\beta+\gamma\right)}}{16N\V{y}^2}
\end{equation*}
Furthermore, by replacing $(\alpha,\beta,\gamma)$ by their expression we find \Equref{Eq:Var_IA}.
\end{proof}
By changing $(\alpha,\beta,\gamma)$ accordingly we establish the variance of $\hat{ST}_{IA}$ as,
\begin{equation*}
\tau_{IA}^2 = \frac{\V{\left(y^A-y^{B_u}\right)^2+\left(y^B-y^{A_u}\right)^2 -ST_u\left(\left(y^A-y^B\right)^2+\left(y^{A_u}-y^{B_u}\right)^2\right)}}{16N\V{y}^2}
\end{equation*}
which is \Equref{Eq:Var_IA_Total}.
\end{appendices}

\section{References}
\bibliographystyle{chicago}

\end{document}